\documentclass[sigconf,nonacm]{acmart}

\setcopyright{none}
\acmDOI{}

\usepackage{tikz}
\usetikzlibrary{shapes.geometric, arrows.meta, positioning, calc}
\usepackage{booktabs}
\usepackage{multirow}

\newcommand{\madaretym}{(\textit{mad\=ar},
  \raisebox{-0.28ex}{\includegraphics[height=1.55ex]{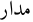}})}

\definecolor{cData}{HTML}{1B7F4B}   
\definecolor{cInstr}{HTML}{1F5FB0}  
\definecolor{cALU}{HTML}{D9690A}    
\definecolor{cXfer}{HTML}{7B3FA0}   
\definecolor{cIO}{HTML}{0E7C86}     
\definecolor{cRail}{HTML}{8C9BA5}   

\newcommand{\phrase}[1]{\emph{#1}}
\newcommand{\rp}{\ensuremath{(\mathit{ring}, \mathit{phase})}}

\begin{document}

\title{MADAR: An Address-Free Processor}

\author{Mohamed Amine Bergach}
\email{mbergach@gmail.com}
\affiliation{%
  \institution{Illumina}
  \city{San Diego}
  \state{California}
  \country{USA}
}
\renewcommand{\shortauthors}{Bergach}

\begin{abstract}
In a modern processor, computing is the cheap part. Most of its area and energy
go to \emph{addressing} --- moving operands to and from a register file and cache,
and running the tags, ports, miss queues, and bypass networks that find a value
where it was left. MADAR deletes that machinery by abolishing the address. All
state circulates in rings of slots that advance one position per clock;
instructions and data ride in the same slots; a value is named by its place in an
orbit --- a \rp{} coordinate --- not by an address; a fixed station computes when
a circulating instruction sweeps past its operands, on a schedule set at compile
time; and a hierarchy of rings of increasing period replaces the cache hierarchy,
movement between them scheduled rather than triggered by a miss. No prior
circulating-store, dataflow, or statically scheduled machine combines all four of
these. We define the execution model, validate it in a cycle-accurate
register-transfer-level implementation, show it \emph{compilable} --- a
constructive scheduler emits programs cross-checked against the implementation ---
and price it with a first-order energy model. The payoff is clearest for AI
acceleration: the multiply--accumulate at the heart of every matmul and
convolution compiles to a streaming form whose energy per operation stays flat as
the reduction grows, and the operand reuse that makes matrix multiplication
efficient is carried by the ring-period hierarchy --- the memory hierarchy doing
by rotation what a cache does by tags. MADAR is a new design point for any
computation whose data movement is known before the program runs.
\end{abstract}

\maketitle

\section{Introduction}
\label{sec:intro}

Most of the machinery in a modern core exists to reconcile a tension: the
datapath streams operands through functional units, but the memory system must
bring every value to rest at a fixed location so it can be found again on demand.
Register files with many ports, caches with their tag arrays and
miss-status registers, memory arbiters, load--store queues, and operand-bypass
networks all exist to stop a value, give it an address, and deliver it again when
an instruction asks for it. This machinery, and not arithmetic, accounts for the
bulk of a core's area and energy. In a detailed energy measurement of a
general-purpose processor running a video encoder, the functional units consumed
under six percent of the total, and more than ninety percent was overhead
relative to the operations actually performed~\cite{Hameed2010}.

MADAR\footnote{MADAR \madaretym: Arabic for ``orbit.''} removes the
contradiction by abolishing stillness. If storage itself circulates, a value
never needs an address: it occupies a position in an orbit and reaches a
functional unit on a schedule fixed when the program is laid out. There is no
fetch, because nothing waits to be fetched, and no cache miss, because nothing is
requested reactively.

Concretely, MADAR is characterized by the conjunction of four properties:
\begin{itemize}
  \item[(a)] \textbf{Circulating storage.} All state is held in rings of slots
    that advance one position per cycle; a value is named by a \rp{} coordinate
    rather than an address. There is no register file, no cache, and no
    random-access memory in the programmer's model.
  \item[(b)] \textbf{Co-circulation.} Instructions and data occupy the same
    slots and circulate together, indistinguishable as storage.
  \item[(c)] \textbf{Collision execution.} A fixed station computes when an
    instruction sweeps past it, taking operands named by their rotational
    offset, on a schedule fixed at compile time --- with no dynamic matching, no
    scoreboard, and no wakeup.
  \item[(d)] \textbf{A period hierarchy for a memory hierarchy.} Rings of
    geometrically increasing period stand in for the levels of a cache, with
    transfer between them scheduled at compile time rather than triggered by
    demand.
\end{itemize}

Circulating storage is not itself new. It is as old as the mercury delay lines of
the first stored-program computers, and it reappears today in racetrack and
recirculating-photonic memories. Temporal operand naming, static instruction
scheduling, and ring-shaped pipelines have each been built many times. What we
believe is new is the combination of all four properties in one general-purpose
machine. Prior systems reach \phrase{pairs} of them but never the full set, and
in particular none uses a hierarchy of ring periods as its memory hierarchy;
Section~\ref{sec:related} places MADAR against these systems in detail.

This paper makes six contributions: it defines the MADAR execution model
precisely enough to implement, including the addressing invariant on which its
correctness turns (Section~\ref{sec:arch}); situates it among the
circulating-store, temporal-addressing, and dataflow architectures it resembles
(Section~\ref{sec:related}); validates the execution semantics in a
cycle-accurate register-transfer-level implementation (Section~\ref{sec:impl});
shows the model is \emph{compilable} rather than hand-seatable, with a
constructive scheduler whose every emitted program is cross-checked against that
implementation (Section~\ref{sec:compiler}); prices the architecture with a
first-order energy model that locates the per-instruction cost at which the
deleted addressing machinery outweighs the rotation it adds
(Section~\ref{sec:eval}); and tests the design point on a streaming-inference
case study, where the operand reuse of dense linear algebra routes through the
period hierarchy (Section~\ref{sec:ai}). An empirical study on real workloads,
and the systems questions a general-purpose machine must answer, remain open
(Section~\ref{sec:future}). MADAR targets computations whose data movement is
known ahead of time, and streaming AI inference is the leading instance: a neural
network layer is multiply--accumulates over a dataflow fixed before the program
runs, exactly the choreography the architecture rewards, and the case study of
Section~\ref{sec:ai} shows it is also where MADAR's efficiency argument pays. We
return to the boundaries of that design point in Section~\ref{sec:discussion}.

\section{The MADAR Architecture}
\label{sec:arch}

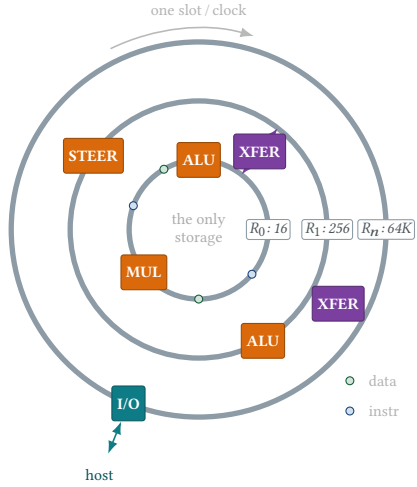
\begin{figure}[t]
  \centering
  \begin{tikzpicture}[
      every node/.style={font=\scriptsize},
      cstn/.style={draw=cALU!70!black, fill=cALU, text=white,
                   font=\scriptsize\bfseries, rounded corners=1pt,
                   inner sep=2pt, minimum height=4.6mm},
      xstn/.style={draw=cXfer!70!black, fill=cXfer, text=white,
                   font=\scriptsize\bfseries, rounded corners=1pt,
                   inner sep=2pt, minimum height=4.6mm},
      iostn/.style={draw=cIO!70!black, fill=cIO, text=white,
                    font=\scriptsize\bfseries, rounded corners=1pt,
                    inner sep=2pt, minimum height=4.6mm},
      rlab/.style={draw=cRail, fill=white, text=cRail!55!black,
                   font=\scriptsize\itshape, inner sep=1.2pt,
                   rounded corners=1pt},
    ]
    \def\Rz{0.92}\def\Ro{1.70}\def\Rt{2.48}
    \draw[cRail, line width=2pt] (0,0) circle (\Rz);
    \draw[cRail, line width=2pt] (0,0) circle (\Ro);
    \draw[cRail, line width=2pt] (0,0) circle (\Rt);
    \node[gray!65, align=center] at (0,0) {the only\\storage};
    \node[rlab] at (0:\Rz) {R$_0$:\,16};
    \node[rlab] at (0:\Ro) {R$_1$:\,256};
    \node[rlab] at (0:\Rt) {R$_n$:\,64K};
    \node[cstn] at (90:\Rz)  {ALU};
    \node[cstn] at (218:\Rz) {MUL};
    \node[cstn] at (145:\Ro) {STEER};
    \node[cstn] at (300:\Ro) {ALU};
    \draw[{Latex[length=1.5mm]}-{Latex[length=1.5mm]}, cXfer, semithick]
      (52:\Rz) -- (52:\Ro);
    \node[xstn] at (52:{(\Rz+\Ro)/2}) {XFER};
    \draw[{Latex[length=1.5mm]}-{Latex[length=1.5mm]}, cXfer, semithick]
      (332:\Ro) -- (332:\Rt);
    \node[xstn] at (332:{(\Ro+\Rt)/2}) {XFER};
    \node[iostn] (io) at (248:\Rt) {I/O};
    \draw[{Latex[length=1.7mm]}-{Latex[length=1.7mm]}, cIO, semithick]
      (io) -- (248:{\Rt+0.72});
    \node[cIO!65!black] at (248:{\Rt+1.02}) {host};
    \foreach \ang/\col in {120/cData, 160/cInstr, 270/cData, 320/cInstr}
      {\node[draw=\col!70!black, fill=\col!20, circle, inner sep=1.05pt]
         at (\ang:\Rz) {};}
    \draw[-{Latex[length=1.8mm]}, gray!55, semithick]
      (116:\Rt+0.18) arc (116:74:\Rt+0.18);
    \node[gray!60] at (90:\Rt+0.42) {one slot\,/\,clock};
    \node[draw=cData!70!black, fill=cData!20, circle, inner sep=1.05pt] at (2.0,-2.0) {};
    \node[anchor=west, gray!60] at (2.14,-2.0) {data};
    \node[draw=cInstr!70!black, fill=cInstr!20, circle, inner sep=1.05pt] at (2.0,-2.4) {};
    \node[anchor=west, gray!60] at (2.14,-2.4) {instr};
  \end{tikzpicture}
  \caption{The MADAR processor as a whole. A hierarchy of rings of increasing
    period (three shown; the design uses more) is the machine's \emph{entire}
    storage, standing in for a cache hierarchy. Compute and control stations
    (orange --- ALU, multiplier, steer) are fixed to the rings and execute
    packets as they sweep past. Transfer stations (purple) bridge adjacent
    rings, promoting a value to a shorter, faster ring or demoting it to a
    longer, slower one on a compile-time schedule --- the role of a cache,
    without misses. A single I/O station (teal) is the only path to the host.
    There is no program counter, no register file, and no cache: the rings
    carry code and data alike, and the period hierarchy \emph{is} the memory
    hierarchy.}
  \label{fig:arch}
\end{figure}

MADAR replaces the two primitives of the von Neumann machine --- a random-access
store and a program counter that walks it --- with a single primitive: rotation.
An address is a name for a value that stays put, chosen so the value can be found
wherever it was left; it is a name for stillness. A \rp{} coordinate is a name
for a value's place in a motion the whole machine shares. Because every value
advances on the same clock, a compiler that fixes the layout of the rotation
knows where every value will be on every cycle, and the question ``where is this
value?'' has an answer fixed at compile time. The model has four elements --- the
rings that hold all state, the packets that fill them, the stations that compute,
and the steering and transfer that provide control flow and a memory hierarchy
--- which we take in turn. The numeric parameters below are representative
working values, not claims about an optimum. Figure~\ref{fig:arch} shows the
machine as a whole; the rest of this section builds it up element by element.

\subsection{Storage: rings}

The machine's only storage is a set of rings $R_0\dots R_n$, each a circular
sequence of slots that advances one slot per clock. Their periods increase
geometrically; representative values are 16, 256, 4K, and 64K slots. A value has
no address, only a \rp{} coordinate: which ring it orbits in, and where in the
orbit it currently sits.

The physical realization is tiered, and it is worth being precise about it,
because the architecture's efficiency argument rests on it. The small, fast rings
are literal register chains --- real shift registers, cheap at that length. The
large, slow rings are \phrase{not} shift registers; they are single-port SRAM
banks read and written through a rotating pointer --- logically circulating,
physically a sequential sweep of one port. That sequential, portless access ---
with no per-access address decode, no tags, no arbitration, and no
multi-porting --- is the cheapest access pattern a RAM offers, and it is what the
architecture trades random access for. The claim is not that storage is free, nor
that the SRAM array disappears; it is that the machinery of random access
\phrase{around} the array is no longer needed.

\subsection{Packets}

Every slot holds exactly one \phrase{packet}. A data packet carries a 64-bit
payload and a small tag; an instruction packet carries an opcode, operand
references, a result destination, and a predicate. Code and data are the same
stuff in the same rotation, distinguished only by a kind field. The slot format
is fixed (Table~\ref{tab:slot}).

\begin{table}[t]
  \caption{Slot format. Every slot is one packet of this shape; $W{=}8$ is the
    operand window, and the ring period $P$ takes one of the representative
    values $\{16, 256, 4\mathrm{K}, 64\mathrm{K}\}$.}
  \label{tab:slot}
  \small
  \begin{tabular}{@{}lll@{}}
    \toprule
    field & width & meaning \\
    \midrule
    \texttt{kind}    & 2\,b  & \texttt{BUBBLE} (empty) / \texttt{DATA} / \texttt{INSTR} \\
    \texttt{op}      & 3\,b  & opcode (\texttt{INSTR}): ADD, SUB, CMPLT, STEER, XFER \\
    \texttt{src\_a}  & 4\,b  & operand-A offset, $1\dots W$ ahead (\texttt{INSTR}) \\
    \texttt{src\_b}  & 4\,b  & operand-B offset, $1\dots W$ ahead (\texttt{INSTR}) \\
    \texttt{dst}     & 4\,b  & result offset ahead (\texttt{STEER}: kill-run start) \\
    \texttt{payload} & 64\,b & \texttt{DATA}: the value. \texttt{STEER}: kill count. \\
    \bottomrule
  \end{tabular}
\end{table}

\subsection{Stations and collision execution}
\label{sec:collision}

Compute stations are fixed hardware mounted at fixed positions on the rings: an
ALU, a multiplier, a steer unit, a transfer unit, an I/O port. A station has no
queue and no scheduler. An instruction packet executes when it sweeps past a
station of its class. Its operands are named not by address but by their
\phrase{rotational offset} --- a position within a window of $W{=}8$ slots ahead,
realized as a small shift-register window --- and its result is injected into a
scheduled slot. There is no dynamic matching, no scoreboard, and no wakeup:
arrival \phrase{is} the schedule. The program is a seating arrangement, and every
latency is fixed and known.

The window $W$ is a hard limit on reach, not a convenience. A producer and a
consumer more than $W$ slots apart on a ring cannot be connected by a single
instruction; the value is relayed forward by a chain of scheduled copies --- each
a same-ring \texttt{XFER} --- until it lands within the consumer's window. The
hardware bypass network of a conventional core, which forwards a result to a
waiting consumer by dynamic match, is thus replaced by a fixed window plus, when
reach is exceeded, explicit compiler-emitted copies. Those copies are not free in
space: each occupies a slot of its own, so long-range dependences trade against a
ring's capacity, and a kernel dense in them may fail to fit.

The one subtlety in the model is how a result is addressed back into the
rotation, and it must be stated precisely. Each cycle, the contents of slot $i$
move to slot $(i{+}1)\bmod P$, so packets pass a station at a fixed position in
order of decreasing initial index. An instruction names a packet by an offset $k$
ahead of itself: the packet at offset $k$ is the one that passed the station $k$
cycles earlier, and when the instruction reaches the station and fires, that
offset resolves to a definite slot. \textbf{All naming is in pre-shift
coordinates: an offset names the \phrase{packet} currently at a given index, not
the register that index denotes.} Mechanically, a write to pre-shift index $j$ is
clocked into register $(j{+}1)\bmod P$ on the same edge as the shift --- which is
exactly where the named packet lands after that shift. The write therefore
replaces the named packet; the register index is an implementation detail, and
the packet is the invariant. Concretely, when an instruction at the station names
a packet currently one position behind it, the result is clocked into the
register that packet moves into on the same edge, overwriting it in place.

\subsection{A worked example: the counted sum}
\label{sec:sumloop}

The simplest program that exercises the whole model is a counted sum. On a ring
of period $P{=}16$ with a single ALU station, five packets are seated as in
Table~\ref{tab:sumloop}: an accumulator, a counter, the constant one, and two
add instructions forming the loop body.

\begin{table}[t]
  \caption{Seating a counted-sum loop on a $P{=}16$ ring with one ALU station.
    Packets pass the station in order of decreasing initial index, so within each
    revolution the $\mathtt{acc}\mathrel{+}=i$ add fires one cycle before the
    $\mathtt{i}\mathrel{+}=1$ add. Both adds read the \phrase{old} $i$; the
    seating order is what makes the loop accumulate $0{+}1{+}\dots{+}(R{-}1)$.}
  \label{tab:sumloop}
  \small
  \begin{tabular}{@{}clll@{}}
    \toprule
    slot & packet & role / effect & fires \\
    \midrule
    5 & \texttt{DATA acc=0} & accumulator & --- \\
    4 & \texttt{DATA i=0}   & counter     & --- \\
    3 & \texttt{DATA one=1} & constant    & --- \\
    2 & \texttt{INSTR ADD a=3 b=2 d=3} & body: $\mathtt{acc}\mathrel{+}=i$ & 14 \\
    1 & \texttt{INSTR ADD a=3 b=2 d=3} & body: $\mathtt{i}\mathrel{+}=1$   & 15 \\
    \bottomrule
  \end{tabular}
\end{table}

The two adds fire one cycle apart within a revolution, and the order matters: the
$\mathtt{acc}\mathrel{+}=i$ add reads \texttt{acc} and $i$ and writes the new
\texttt{acc}; the $\mathtt{i}\mathrel{+}=1$ add reads the \phrase{same} $i$ ---
not yet incremented, since $i$ is written only by that very instruction --- with
the constant, and writes the new $i$. The seating order, with
the accumulator update ahead of the counter update, is what makes the loop
compute $0{+}1{+}\dots{+}(R{-}1)$ rather than $1{+}2{+}\dots{+}R$. Nothing is
fetched: the body is parked in the ring and re-executes simply because it keeps
coming back around.

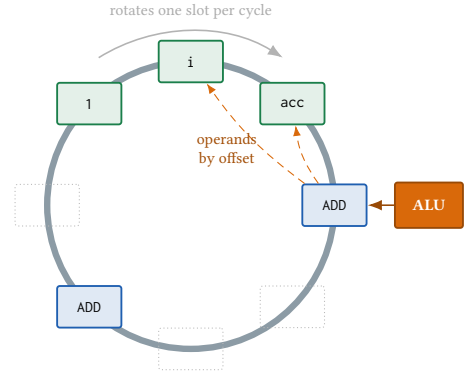
\begin{figure}[t]
  \centering
  \begin{tikzpicture}[
      every node/.style={font=\scriptsize},
      slot/.style={draw=cData, line width=0.7pt, rounded corners=1pt,
                   fill=cData!12, minimum width=8.5mm, minimum height=5.5mm,
                   inner sep=1pt},
      instr/.style={draw=cInstr, line width=0.7pt, rounded corners=1pt,
                    fill=cInstr!14, minimum width=8.5mm, minimum height=5.5mm,
                    inner sep=1pt},
      mt/.style={draw=gray!55, densely dotted, rounded corners=1pt,
                 minimum width=8.5mm, minimum height=5.5mm, inner sep=1pt},
      stn/.style={draw=cALU!70!black, line width=0.7pt, fill=cALU, text=white,
                  font=\scriptsize\bfseries, rounded corners=1pt,
                  minimum width=9mm, minimum height=6mm},
    ]
    \def\R{1.9}
    \draw[cRail, line width=2.4pt] (0,0) circle (\R);
    \draw[-{Latex[length=2mm]}, gray!60, semithick]
      (122:\R+0.40) arc (122:58:\R+0.40);
    \node[gray!65] at (90:\R+0.66) {rotates one slot per cycle};
    \node[instr] (add1) at (0:\R)   {\texttt{ADD}};
    \node[slot]  (acc)  at (45:\R)  {\texttt{acc}};
    \node[slot]  (ival) at (90:\R)  {\texttt{i}};
    \node[slot]  (one)  at (135:\R) {\texttt{1}};
    \node[mt]    (e1)   at (180:\R) {};
    \node[instr] (add2) at (225:\R) {\texttt{ADD}};
    \node[mt]    (e2)   at (270:\R) {};
    \node[mt]    (e3)   at (315:\R) {};
    \node[stn] (alu) at (0:\R+1.25) {ALU};
    \draw[-{Latex[length=1.9mm]}, cALU!70!black, semithick] (alu) -- (add1);
    \draw[-{Latex[length=1.4mm]}, densely dashed, cALU]
      (add1) to[bend left=14] (acc);
    \draw[-{Latex[length=1.4mm]}, densely dashed, cALU]
      (add1) to[bend left=9] (ival);
    \node[cALU!75!black, align=center] at (57:\R*0.46) {operands\\by offset};
  \end{tikzpicture}
  \caption{A single ring, drawn with eight slots, holding the counted-sum loop of
    Section~\ref{sec:sumloop}. Each slot carries one packet --- an instruction
    (blue), a data value (green), or nothing (grey, dotted) --- and every slot
    advances one position per cycle past stations fixed to the ring. The ALU
    (orange) executes whichever instruction is passing it, reading its operands from
    nearby slots by rotational offset; here the add reads \texttt{acc} and
    \texttt{i}. Other stations, such as steer and transfer, occupy other fixed
    positions. Storage, code, and execution share one rotation; there is no
    separate memory, register file, or fetch unit.}
  \label{fig:ring}
\end{figure}

After $R$ revolutions --- $R{\cdot}P$ cycles, with every packet home at its seed
slot --- the counter holds $R$ and the accumulator holds
\[
  \mathtt{acc} = 0 + 1 + \dots + (R{-}1) = \frac{R(R{-}1)}{2}.
\]
For $R{=}10$ this is $45$, and for $R{=}15$ it is $105$. A loop iteration is
precisely one revolution of the ring. Figure~\ref{fig:ring} zooms into a single
ring with this seating and its ALU station.

\subsection{Control flow: steering, not branching}

With no program counter, there is nothing to redirect. Control flow becomes a
question of which already-circulating packets stay alive. A steer station
evaluates a predicate and either removes an instruction group --- turning its
slots to bubbles, so it never executes again --- or leaves it in place. A short
forward branch computes both arms and selects the survivor. A loop runs once per
revolution while its body remains seated, and it ends when a steer station, armed
by a comparison computed elsewhere on the ring, removes that body as it passes;
the surviving packet is the result, which a transfer carries to an outer ring.
The counted sum of Section~\ref{sec:sumloop} terminates exactly this way: a
comparison sets a done flag, and a steer keyed on that flag clears the body on the
revolution the flag turns true.

\subsection{The memory hierarchy: scheduled transfer}
\label{sec:transfer}

Transfer stations sit where rings meet and move packets between them at scheduled
phases. Promotion to a shorter-period ring plays the role of caching, demotion to
a longer one the role of eviction --- neither reactive, with no misses. The cost
of reaching a value is
its phase distance to the station that needs it: on a 16-slot ring, at most
fifteen cycles; on a 64K-slot ring, up to 64K, and on average half that. The
period hierarchy is therefore a latency hierarchy, and laying a kernel across it
--- promoting what is needed soon to a short ring, demoting what is needed later
to a long one --- is the compiler's central task. Ring capacity
is likewise a static property: a kernel whose live packets exceed a ring's period
is placed on the next-longer ring, and a program that fits no ring does not
compile. There is no dynamic spill.

The one genuinely hard case is data whose location is not known until run time ---
a pointer whose target ring and phase the compiler cannot pre-place. Such a value
is reached through a \phrase{rendezvous} station, which holds a single requested
phase while the ring carries one free-running phase counter; when the counter
matches, the station captures or overwrites the slot then passing. The match is
one comparator against one counter, not an associative search across slots, so
the mechanism does not reintroduce the tag arrays and ports that random access
would. The cost is paid in latency: the value is, on average, half a rotation
away. This is the price of an access the schedule cannot anticipate, and it bounds
the workloads to which the architecture is suited (Section~\ref{sec:discussion}).

\section{Design Rationale}
\label{sec:rationale}

MADAR is built to spend as little as possible on \phrase{addressing} --- moving
operands to and from a random-access store, the term that dominates a
conventional core's energy (Section~\ref{sec:intro}). Three arguments support the
design, each a claim the cost model of Section~\ref{sec:eval} is built to test.

First, the architecture deletes the addressing machinery rather than optimizing
it. If storage circulates and operands arrive by rotation, a slot's identity is
its phase, so there are no tags; one packet occupies one slot per cycle, so there
is no arbitration; the only memory port is the single sequential SRAM port, swept
by a rotating pointer with no random address decode and no port replication;
nothing is requested reactively, so there are no miss-status registers; and the
operand window takes the place of the bypass network. The structures a
conventional core spends most of its energy on are absent by construction, and
the cost of ``storage'' collapses toward the cost of the SRAM array itself.

Second, timing is exact rather than best-effort. Because nothing in the memory
system is reactive, every operand's arrival cycle is fixed at compile time, and
the emitted schedule is the executed schedule, cycle for cycle. This distinguishes
MADAR from statically scheduled VLIW, whose bundles issue against a random-access
register file and reactive caches, where a miss invalidates the schedule and
forces the stall-and-replay logic the compiler cannot model. MADAR has no
reactive memory to invalidate the plan.

Third, loops cost nothing to supply. A parked loop body re-executes once per
revolution with no fetch, no decode, and no branch prediction --- which matters
because instruction supply, not arithmetic, is roughly a third of core
energy~\cite{Hameed2010}, and a machine that supplies a loop body for free
attacks that term directly.

\section{Related Work}
\label{sec:related}

Circulation, temporal addressing, static scheduling, and ring-shaped topologies
have each been realized many times. The recurring pattern is that prior art
reaches \phrase{pairs} of MADAR's four properties but never the full set. Two
boundaries matter most --- the Mill belt on operand naming and the Groq processor
on scheduling --- alongside the delay-line ancestors that own circulation itself.

\paragraph{The delay-line ancestors.} In the Pilot ACE (1950) and its production
successor DEUCE (1955), all storage --- code and data alike --- lived in
circulating mercury delay lines; each instruction named a (line, wait, timing)
coordinate, and ``optimum programming'' placed the next instruction at the phase
where it was just emerging from a line, a literal \rp{} discipline with no program
counter~\cite{PilotACE}. The same shape appears in the EDSAC/EDVAC
lineage~\cite{Wilkes1951}, in the IBM~650 with SOAP optimal drum
coding~\cite{IBM650}, in the Bendix~G-15 with its two tiers of recirculating lines
and accumulators that were themselves short lines~\cite{BendixG15}, and in the
Datapoint~2200, which used MOS shift registers as primary store~\cite{Datapoint2200}.
These machines own circulating, co-circulating storage --- properties (a) and (b)
--- but they compute at a small \phrase{central} arithmetic unit, not at stations
distributed along the circulation, and their latency tiers were bridged by
hand-coded placement rather than an architectural transfer mechanism. They are the
genuine ancestors of co-circulation; MADAR's departure from them is distributed
collision execution and a compiler-scheduled period hierarchy.

\paragraph{The Mill belt.} The Mill names operands by their temporal production
rank on a fixed-length queue, statically and without register write-back, the
closest prior art to MADAR's offset-based operand naming~\cite{MillBelt, MillDocs}.
But the belt is a \phrase{logical} list of the last several results --- the patent
allows it to be a circular buffer, a register file, or a content-addressable
memory --- and conventional fetch and caches sit behind it. MADAR's \rp{} is a
\phrase{geometric} coordinate on a \phrase{physically} circulating medium whose
period is a design parameter, naming \phrase{all} storage with instructions in the
same slots. The distinction is precise: a geometric circulation address over all
storage, versus a logical recency rank over transient results.

\paragraph{Phase-scheduled loop hardware.} The HP ShiftQ accelerator routes loop
results through a shift-cell chain clocked at the loop initiation interval, with
operands named by phase offset and the loop running once per
revolution~\cite{HPShiftQ} --- the nearest prior echo of iteration-as-revolution.
But it is a linear chain with no wraparound and no \rp{} coordinate, its control
is hardwired rather than carried as instructions in the medium, and it has a
single period. It is a fixed-function loop accelerator, not a general-purpose
machine.

\paragraph{Ring-dataflow and dataflow machines.} Packets circulate a closed
pipeline through fixed stations, sometimes carrying an instruction field with
their operands, in Sharp's circular pipeline~\cite{SharpCircular}, Mitsubishi's
data-driven processor~\cite{MitsubishiDataflow}, the Manchester prototype dataflow
machine and its multi-ring extension~\cite{Gurd1985, Barahona1986}, and the broader
dataflow tradition from Dennis~\cite{Dennis1974} through Monsoon's explicit token
store~\cite{Papadopoulos1990}. Every one of them fires by \phrase{dynamic tag
matching} --- a matching unit pairs tokens at run time --- which is the precise
opposite of MADAR's compile-time collision, and their instructions are generally
held in a separate store. The closest co-circulation in the literature is an
asynchronous delay-line ring in which opcode and operands travel
together~\cite{AsyncDelayRing}, but it fires on demand by latch-vacancy detection,
without a shift clock and so without static timing, and it draws on a separate
function store. The spatial dual --- instructions fixed in place with data moving
to them --- is WaveScalar and the EDGE/TRIPS line~\cite{Swanson2003, Burger2004},
which inverts MADAR's data-fixed, instruction-moving discipline and still matches
dynamically.

\paragraph{Modern static scheduling.} The Groq processor performs compiler
choreography with cycle-exact timing and no misses at production
scale~\cite{Groq2020, GroqPatent}, the strongest contemporary instance of MADAR's
scheduling discipline. But its storage is \phrase{stationary} SRAM with separate
instruction queues; data streams across a mesh rather than circulating; and there
is one flat memory level. The compiler-exact operand transport of
RAW~\cite{Taylor2002} and the compiler-managed stream register file of
Imagine~\cite{Kapasi2002} capture the spirit of scheduled transfer in place of
caching, but again over flat, stationary memory. The software-pipelining and
modulo-scheduling techniques that make such schedules possible rotate an address
\phrase{map} per iteration~\cite{Lam1988, Rau1994}; in MADAR the storage medium
itself rotates, with the code in it.

\paragraph{Other rotations and circulating media.} Several systems rotate
something other than storage. Systolic and wavefront arrays clock data through
fixed-function elements without a circulating stored program~\cite{Kung1978}. The
counterflow pipeline matches instructions and results
dynamically~\cite{Sproull1994}; its circular Intel descendants ring-wrap that
bypass network inside a conventional out-of-order core~\cite{IntelCounterflowA,
IntelCounterflowB}. Barrel processors
rotate \phrase{contexts}, not storage: the CDC~6600 peripheral processor presented
ten register sets round-robin to one pipeline~\cite{Thornton1970}, as
multithreaded cores do today. Circulating-memory technologies circulate storage
beneath a conventional ALU --- magnetic-bubble loop units~\cite{BubbleLoopALU},
processing-in-racetrack at fixed heads~\cite{Ollivier2022}, reconfigurable arrays
fed from rotating buffers~\cite{WaveNicolCGRA}, tapped delay-line
buses~\cite{KoneskyTapped}, marching memories~\cite{MarchingMemory},
circular-buffer ISA operands~\cite{CircBufOperand}, and instruction-cycling over
conventional memory~\cite{AutonCycling} --- but none carries instructions in the
medium with a period hierarchy. Recirculating delay-line memory now runs at tens
of gigahertz~\cite{Volk2023}, feasible at speed but with no execution in the loop.
Distinct from all of these, a parallel \phrase{temporal-computing} line --- race
logic and its successors --- encodes a value as a signal's arrival time and
computes by the order in which wavefronts race~\cite{RaceLogic,
TemporalStateMachines}: it computes \phrase{in} time where MADAR \phrase{addresses}
in time, with no circulating store, no co-circulating instructions, and no period
hierarchy. The two share only the medium of time.

\begin{table*}[t]
  \caption{Representative prior architectures against MADAR's four defining
    properties: (a) all storage in circulating rings with \rp{} addressing;
    (b) instructions and data co-circulating in the same slots; (c) static
    collision execution at fixed stations with rotation-offset operands; (d) a
    hierarchy of ring periods serving as the memory hierarchy. No system outside
    MADAR carries more than two non-\textsc{no} marks, and none realizes (d) as
    an architectural mechanism --- DEUCE's hand-coded latency tiers are the only
    partial.}
  \label{tab:matrix}
  \small
  \begin{tabular}{@{}lcccc@{}}
    \toprule
    System & (a) rings, \rp{} & (b) co-circulate & (c) static collision & (d) period hierarchy \\
    \midrule
    \textbf{MADAR} & \textbf{yes}$^{\dagger}$ & \textbf{yes} & \textbf{yes} & \textbf{yes} \\
    Pilot ACE / DEUCE~\cite{PilotACE}        & yes     & yes     & no      & partial (hand-coded) \\
    Bendix G-15 / IBM 650+SOAP~\cite{BendixG15, IBM650} & yes & yes & no & no \\
    HP ShiftQ~\cite{HPShiftQ}                & partial & no      & partial & no \\
    Groq TSP~\cite{Groq2020}                 & no      & no      & partial & no \\
    RAW scalar operand net~\cite{Taylor2002} & no      & no      & partial & no \\
    Wave CGRA~\cite{WaveNicolCGRA}           & no      & partial & partial & no \\
    Async delay-line ring~\cite{AsyncDelayRing} & partial & partial & no   & no \\
    Sharp circular pipeline~\cite{SharpCircular} & partial & partial & no  & no \\
    Mitsubishi data-driven~\cite{MitsubishiDataflow} & partial & partial & no & no \\
    Manchester dataflow~\cite{Gurd1985}      & no      & no      & no      & no \\
    Mill belt~\cite{MillBelt}                & no      & no      & partial & no \\
    Counterflow pipeline~\cite{Sproull1994}  & partial & no      & no      & no \\
    \bottomrule
  \end{tabular}
  \par\smallskip
  \raggedright\footnotesize $^{\dagger}$\,Physically literal on the short
  shift-register rings; logical --- a rotating-pointer sweep over single-port
  SRAM --- on the large tiers, where the programmer's model is identical.
\end{table*}

The comparison makes the pattern visible. Co-circulating storage, properties (a)
and (b), exists only in the delay-line machines with their central arithmetic
units; static collision, property (c), exists in modern scheduling work but over
stationary storage; and column (d) is empty outside MADAR but for DEUCE's
hand-coded tiers. No prior system combines all four, and the sharpest of them is
the last --- a ring-period hierarchy used \phrase{as} the memory hierarchy, found
nowhere else as an architectural mechanism.

One qualification carries the weight of the novelty and must be made operational.
Property (a) is physically literal only on the short shift-register rings; the
large tiers are single-port SRAM swept by a rotating pointer
(Table~\ref{tab:matrix},~$\dagger$). What then separates those tiers from a
stationary-SRAM scheduler such as Groq is not rhetoric but mechanism: no address
is presented to the array --- only a free-running phase counter and a rotational
offset resolved by the schedule, with no decoder and no tag --- where Groq issues
explicitly addressed reads from stationary banks into stream queues. The medium
changes across tiers; the addressing discipline does not, and it is that
discipline --- neither ``compute in circulating memory,'' which is old, nor
``static scheduling,'' which Groq ships --- that is the architecture's novel core.

\section{Implementation and Validation}
\label{sec:impl}

We implemented MADAR as a cycle-accurate register-transfer-level model in
SystemVerilog --- a ring with seeded contents and write ports, and ALU, steer,
and transfer stations --- and simulated it with Verilator.\footnote{The full
artifact --- RTL, the cross-checked functional model, the scheduling compiler, the
energy model, and the AI demonstrator --- is open source under the MIT license at
\url{https://github.com/aminems/Madar}.} A suite of directed
tests exercises each mechanism of the execution model and confirms it behaves as
specified (Table~\ref{tab:tests}); all tests pass. The purpose of this exercise is
to establish that the mechanics are sound, not to evaluate performance: no
benchmark, instruction-throughput figure, or energy number is claimed here ---
those follow only once programs are \emph{compiled} to the model
(Section~\ref{sec:compiler}) and priced (Section~\ref{sec:eval}).

\begin{table}[t]
  \caption{Behaviors validated in the cycle-accurate implementation. Each row
    exercises one mechanism of the execution model; the outcome is the directed
    test's checked result.}
  \label{tab:tests}
  \small
  \begin{tabular}{@{}p{0.40\columnwidth}p{0.45\columnwidth}@{}}
    \toprule
    behavior & checked outcome \\
    \midrule
    Rotation is storage: ring contents return to their seating after one full
      period & identity after $P$ cycles, no stray packets \\[2pt]
    Collision execution: an add meets its operands at a fixed station by
      rotation offset & $7{+}35{=}42$, result replaces the named packet \\[2pt]
    Iteration is revolution: the parked sum loop runs once per turn with no
      fetches & $\mathtt{acc}{=}45$ at 10 turns, $105$ at 15 \\[2pt]
    Steering replaces branching: a comparison-armed steer ends the loop & one
      surviving packet, value $45$ \\[2pt]
    Transfer replaces caching: a packet migrates to a longer ring at a scheduled
      phase & stable \rp{} over a 64-cycle orbit \\
    \bottomrule
  \end{tabular}
\end{table}

The validated behaviors correspond one-to-one to the model's claims. Rotation is
storage: a ring's contents return to their seating after exactly one period, with
no packet lost or duplicated. Collision is execution: an add instruction meets its
operands at a fixed station by rotation offset, and its result replaces the named
packet, with no fetch and no register file. Iteration is revolution: the parked
sum loop of Section~\ref{sec:sumloop} runs unchanged, reaching an accumulator of
45 after ten revolutions and 105 after fifteen, the body still seated and
instruction fetches at zero. Steering replaces branching: a predicate computed by
a comparison arms a steer station that removes the loop body on the revolution the
predicate turns true, leaving the single result packet alive. And transfer
replaces caching: a packet migrates from a short ring to a longer one at a
scheduled phase and keeps a stable \rp{} coordinate over a full orbit of the
larger ring.

The directed tests above seat their programs by hand. Two further cross-checked
programs complete the mechanics: a same-ring copy relay (the primitive used to
bridge a long-reach dependence) and a \emph{move} --- an inter-ring transfer
paired with a steer that clears the source, the eviction half of the memory
hierarchy that the transfer and steer tests had shown only separately. Whether
the model can be \emph{compiled to}, rather than hand-seated, is the question
addressed next (Section~\ref{sec:compiler}); the programs the scheduler emits are
validated by this same cross-check against the register-transfer-level model. The
efficiency question is taken up quantitatively in Section~\ref{sec:eval};
contention and port budgets at scale, and a study on real workloads, remain open
(Section~\ref{sec:future}).

\section{The Scheduling Compiler}
\label{sec:compiler}

A model that only a person can seat is a curiosity; a model a compiler can target
is an architecture, and which of the two MADAR is turns on whether a seating
arrangement and a transfer schedule can be \emph{computed} for a kernel from its
dataflow alone --- placing values across the ring hierarchy, inserting copy
relays where reach is exceeded, and meeting ring-capacity bounds. This section
answers that question constructively.

The scheduler takes a kernel as a dataflow graph --- a straight-line computation
or a counted loop --- issues its operations in dependency order at a single
station, and lays the operation and value packets on the ring within the
$W{=}8$-slot operand window. The placement is constructive rather than
exhaustive: a backtracking placer that enumerates seatings becomes intractable
beyond a handful of packets, whereas the constructive scheduler places the
kernels below in well under a second each. Every candidate it forms is
validated by running it on the functional model and comparing against the
kernel's reference result, so a returned program is correct by construction and
check, not by trust in the placement arithmetic. Three mechanisms give it the
reach a real kernel needs.

\paragraph{Copy relays.} When a producer and a consumer fall more than $W$ slots
apart on a ring, no single instruction can connect them
(Section~\ref{sec:collision}). The scheduler keeps each live value in reach by
republishing it: the moment a value's freshest copy would pass out of the window,
it emits a same-ring \texttt{XFER} that copies the value forward into a fresh
slot and redirects later reads to that copy. A long dependence chain, or a value
fanned out to many consumers, is carried by a chain of such copies. A relay is
itself a packet occupying a slot, so a kernel dense in long-range dependences
spends ring capacity on its relays; the cost is real and is priced in
Section~\ref{sec:eval}. The relay span is auto-tuned --- the fewest relays that
admit a seating --- so a kernel that needs none pays nothing.

\paragraph{Placement across the hierarchy.} A kernel given as a sequence of
stages is placed across rings of increasing period, one stage per ring, with each
intermediate carried to the next ring by a scheduled inter-ring \texttt{XFER}.
The transfer is not left to chance. A copy that fired every revolution while the
destination ring rotated would scatter the value across the ring; instead the run
plan choreographs it --- each stage settles on its ring while the others are
held, the cross-ring transfer fires in a single shared advance of both rings, and
the next stage then settles --- so a value crosses from one ring to the next
exactly once, at a phase the schedule fixes. Because that single transfer advance
shifts the destination ring by one slot, each transfer station is placed where
its instruction actually sits when it fires --- which is what makes three-ring
pipelines (and deeper) land correctly, not just two-ring ones. This is a cache
fill performed by the schedule in place of a miss.

\paragraph{Capacity and demotion.} Ring capacity is a static property: a kernel
whose packets --- values, instructions, and relays together --- exceed a ring's
period cannot be seated on it. The scheduler then demotes the kernel to the
next-longer ring in the period hierarchy and retries. A kernel that fits no ring
fails to compile; there is no dynamic spill.

\paragraph{Scope.} The scheduler places straight-line dependence chains and
counted loops, in which each operation consumes at most one previously produced
value together with seeded inputs --- the shape of multiply--accumulate,
polynomial evaluation, and the loops and chains below. A reduction \emph{tree},
in which one operation consumes two separately produced results that must be
co-located within a single window, and a liveness-aware reordering that would
keep more intermediates in reach at once, are not yet handled: both are placement
problems the constructive method can be extended to, and we name them as future
work rather than claim them. Within its scope the scheduler is exact --- every
program it emits is the one validated against the model.

\begin{table}[t]
  \caption{Kernels the scheduler places, each compiled from its dataflow with no
    hand placement, run on the functional model to its reference result, and
    emitted as a register-transfer-level testbench cross-checked slot-for-slot
    against the Verilator model of Section~\ref{sec:impl}. All agree. ``R'' is the
    relay count; the pipeline kernel is placed across two rings with one scheduled
    inter-ring transfer; the fan-out kernel, declared on a ring too small, is
    demoted to a longer one.}
  \label{tab:compile}
  \small
  \begin{tabular}{@{}llcl@{}}
    \toprule
    kernel & ring(s) & R & mechanism exercised \\
    \midrule
    multiply--accumulate        & 16        & 0 & single ring, within window \\
    degree-2 polynomial         & 32        & 0 & single ring, dependence chain \\
    counted sum (10 iters)      & 16        & 0 & parked loop, in-place update \\
    FIR tap (4 iters)           & 16        & 0 & loop with multiply \\
    degree-5 polynomial         & 64        & 3 & relays the variable read each step \\
    12-deep accumulation        & 64        & 5 & relays the constant read each step \\
    two-stage pipeline          & 16, 64    & --& inter-ring transfer (property d) \\
    fan-out (8-wide)            & $16{\to}64$& 6 & capacity demotion \\
    \bottomrule
  \end{tabular}
\end{table}

Table~\ref{tab:compile} lists the kernels the scheduler places and the mechanism
each exercises. The decisive check is the last column of the validation, not
shown in the table because it is uniform: each compiled program is emitted as a
register-transfer-level testbench whose final ring state is compared
slot-for-slot against the same cycle-accurate model that Section~\ref{sec:impl}
validates, and all agree. The polynomial and accumulation kernels require relays
because their variable or constant is read on every step and must be carried along
the ring; the two-stage kernel is placed across two rings of different period
with one scheduled transfer; the fan-out kernel, declared on a 16-slot ring that
cannot hold it, is demoted to a 64-slot ring. The scheduler does not claim to find
the minimal-energy seating --- it computes \emph{a} correct one --- but it settles
the question that separates a machine from a curiosity: MADAR's programs are
computed, not hand-laid.

\section{Cost Model and Evaluation}
\label{sec:eval}

The design rationale of Section~\ref{sec:rationale} is an argument that deleting
the addressing machinery saves more than rotation costs. With a compiler that
places real kernels, the argument can be priced. We build a first-order energy
model: an activity count over a compiled program, scaled by published
per-operation energies, set against an in-order baseline of instruction count
times energy per instruction. The model is deliberately transparent --- its
conclusion is a crossover, not an absolute joule figure --- and it is built to
name the regimes where MADAR loses.

\paragraph{Energies.} The per-event figures are Horowitz's 45\,nm
values~\cite{Horowitz2014}: a 64-bit add at 0.2\,pJ, a multiply at 12.4\,pJ, an
8\,KB SRAM access at 10\,pJ. The pivot is the cost of advancing one slot, priced
three ways: a conservative register-file proxy (6\,pJ, an upper bound on a true
shift), a realistic flip-flop toggle (1\,pJ), and the SRAM access (10\,pJ) for the
large rings swept by a rotating pointer. A copy relay or inter-ring transfer is
one word move at the chosen rate. The baseline is a simple in-order core at
70\,pJ per instruction --- Horowitz's figure for a programmable processor's
per-instruction overhead, consistent with measured embedded cores.

\paragraph{Activity.} For a compiled program the activity is exact rather than
estimated, and rotation is \phrase{clock-gated}: on a shift-register ring only the
slots whose contents change dissipate a word-shift each cycle --- the occupied
slots and the leading edge of each empty run, the bubbles gated off --- while a
large SRAM ring advances its rotating pointer at one read and one write per cycle.
Each instruction fires once per revolution of its ring. The MADAR energy is the
sum of rotation (gated slot-shifts at the per-slot cost), compute (adds and
multiplies at their energies), and relays (transfers at the per-word cost). The
crossover is the baseline energy per instruction at which the two machines meet,
$E_\text{MADAR}/N_\text{instr}$ --- directly comparable to the 70\,pJ figure.

\paragraph{Results.} Table~\ref{tab:energy} reports each kernel's crossover at the
ring the compiler actually lands it on, each ring priced at its realization ---
short rings as shift registers, longer ones as the SRAM rotating pointer. With
idle slots gated, the multiply--accumulate and the two loop kernels --- a few
packets parked on a 16-slot ring --- cross at 26--63\,pJ per instruction, below a
70\,pJ in-order instruction and with no appeal to an undersized ring the hardware
cannot allocate. (The conservative register-file shift proxy raises these; it is
an upper bound we report but do not lean on.) The polynomial and accumulation
chains, pushed onto 32- to 64-slot rings by their packet count and, for the
longer chains, copy relays, cross at 126--205\,pJ and lose, because even gated,
shifting a larger ring's live packets across the settle dominates a small
computation. The reading is the one the thesis predicts: MADAR is an efficiency
win where the ring is sized to the work and a loss where it is not --- a new
design point, not universal supremacy, and one an energy-aware placer would push
further by right-sizing each ring (Section~\ref{sec:future}).

\begin{table}[t]
  \caption{Energy crossover per kernel: the baseline cost-per-instruction at which
    the in-order core and MADAR consume equal energy, with rotation clock-gated
    and each ring at the period the compiler lands it on, priced at its
    realization (short rings as shift registers, longer as the SRAM pointer).
    Below the 70\,pJ in-order baseline, MADAR wins. The loop and
    multiply--accumulate kernels win on their short rings; the polynomial and
    accumulation chains, pushed onto longer rings, lose to rotation.}
  \label{tab:energy}
  \small
  \begin{tabular}{@{}lccr@{}}
    \toprule
    kernel & relays & crossover (pJ/instr) & vs.\ 70\,pJ \\
    \midrule
    counted sum            & 0 & 26  & \textbf{wins} \\
    FIR tap                & 0 & 30  & \textbf{wins} \\
    multiply--accumulate   & 0 & 63  & \textbf{wins} \\
    degree-2 polynomial    & 0 & 126 & loses \\
    degree-5 polynomial    & 3 & 161 & loses \\
    12-deep accumulation   & 5 & 205 & loses \\
    \bottomrule
  \end{tabular}
\end{table}

\paragraph{The reach--capacity tradeoff.} Copy relays are not free: each is a
packet, so long-range dependences trade against capacity. Holding the ring period
at 64 and growing a dependence chain, the relay count climbs with length --- a
chain of 16 operations needs 7 relays --- until the packets overflow the ring and
the kernel must demote: a chain of 24, at 72 packets, no longer fits a 64-slot
ring. Reach and capacity are two sides of one budget, and the compiler spends one
for the other.

\paragraph{Threats to validity.} The model is first-order. Its gated rotation is
an idealized clock-gate (it counts the slots whose contents change, not a measured
flop-enable network); it prices a rotation by one per-word figure rather than a
synthesized netlist; and the baseline is a single published per-instruction
number, not a matched implementation measured on the same flow. The figures are stand-ins, which is why
the result we rest on is the crossover --- robust to the exact magnitudes --- and
not an absolute energy. A synthesis-based comparison of both machines through one
flow is the evaluation this model motivates and does not replace.

\section{A Case Study: Streaming AI Inference}
\label{sec:ai}

The applicability claim --- that MADAR fits computations whose data movement is
fixed in advance (Section~\ref{sec:discussion}) --- deserves a concrete test. The atom of dense inference is
the multiply--accumulate, and a layer is multiply--accumulates tiled and reduced.
We built the inference primitives on MADAR --- a dot product, a streaming inner
product, and a matmul tile --- compiled and validated them against the
cycle-accurate model of Section~\ref{sec:impl}, and priced them with the cost
model of Section~\ref{sec:eval}. The exercise confirms the design point and
sharpens it: the win is real where the ring is sized to the work, a flat
reduction loses exactly as the polynomial chains did, and operand reuse --- the
heart of matrix-multiply efficiency --- turns out to route through the period
hierarchy.

\paragraph{A reduction must be sequential to seat.} A dot product
$\sum_i a_i b_i$ is a reduction, and a flat reduction tree does not seat under the
constructive scheduler: its products are produced independently, are placed with
free choice and scatter, and no add reaches two of them within the $W{=}8$ window
(Section~\ref{sec:compiler}). Written instead as a sequential accumulation
$s_i = s_{i-1} + a_i b_i$, each add consumes the running sum and one fresh product
--- the two most recent results --- and the scheduler seats it with no relays.
Matrix--vector products and matmul tiles follow as compositions of these inner
products, each validated against the model. The same window that bounds the
scheduler thus dictates the dataflow: accumulate, do not tree.

\paragraph{The flat reduction loses; the sized loop wins.} A flat dot product's
crossover climbs with length --- 59, 81, 84, 111\,pJ per instruction at
$N{=}2,4,8,12$ --- and falls below the 70\,pJ in-order baseline only at the
smallest size. Clock-gating does not rescue it, because nothing is idle: a
length-$N$ reduction is $2N$ operations but carries some $6N$ \phrase{live}
packets --- the operands, the products, the partial sums --- so its ring is sized
to $6N$ and every one of those packets shifts each cycle, three times the rotation
per useful operation and growing with $N$. The multiply--accumulate \emph{loop} of
Table~\ref{tab:energy}, by contrast, wins at 26--30\,pJ: a handful of packets on a
16-slot ring, re-executed once per revolution, every revolution doing useful work.
A flat unroll is the wrong shape; a sized loop the right one.

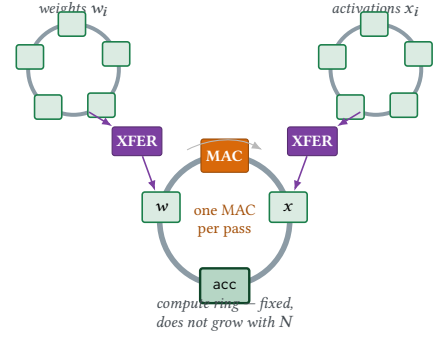
\begin{figure}[t]
  \centering
  \begin{tikzpicture}[
      every node/.style={font=\scriptsize},
      cstn/.style={draw=cALU!70!black, fill=cALU, text=white,
                   font=\scriptsize\bfseries, rounded corners=1pt,
                   inner sep=2pt, minimum height=4.4mm},
      xstn/.style={draw=cXfer!70!black, fill=cXfer, text=white,
                   font=\scriptsize\bfseries, rounded corners=1pt,
                   inner sep=1.6pt, minimum height=3.8mm},
      dslot/.style={draw=cData, fill=cData!16, line width=0.6pt,
                    rounded corners=1pt, minimum width=3.6mm,
                    minimum height=3.2mm, inner sep=0.4pt},
      accs/.style={draw=cData!55!black, fill=cData!32, line width=0.9pt,
                   rounded corners=1pt, minimum width=6.5mm, minimum height=4.6mm},
      bigslot/.style={draw=cData, fill=cData!16, line width=0.6pt,
                      rounded corners=1pt, minimum width=5mm,
                      minimum height=4mm, inner sep=0.5pt},
      rlab/.style={font=\scriptsize\itshape, text=cRail!50!black},
    ]
    \def\Rs{0.6}
    \foreach \cx/\lab in {{-2.0}/{weights $w_i$}, {2.0}/{activations $x_i$}} {
      \draw[cRail, line width=1.5pt] (\cx,1.45) circle (\Rs);
      \node[rlab] at (\cx,2.24) {\lab};
      \foreach \a in {20,92,164,236,308} {
        \node[dslot] at ($(\cx,1.45)+(\a:\Rs)$) {};
      }
    }
    \def\Rc{0.86}
    \draw[cRail, line width=2.1pt] (0,-0.55) circle (\Rc);
    \node[cstn] at ($(0,-0.55)+(90:\Rc)$)  {MAC};
    \node[accs] at ($(0,-0.55)+(270:\Rc)$) {\texttt{acc}};
    \node[bigslot] (wl) at ($(0,-0.55)+(168:\Rc)$) {$w$};
    \node[bigslot] (xl) at ($(0,-0.55)+(12:\Rc)$)  {$x$};
    \node[cALU!75!black, align=center] at (0,-0.55) {one MAC\\per pass};
    \node[rlab, align=center] at (0,-1.78) {compute ring --- fixed,\\does not grow with $N$};
    \draw[-{Latex[length=1.4mm]}, gray!55]
      (0,-0.55) ++(118:\Rc+0.2) arc (118:62:\Rc+0.2);
    \node[xstn] (xw) at (-1.16,0.5) {XFER};
    \node[xstn] (xx) at ( 1.16,0.5) {XFER};
    \draw[-{Latex[length=1.4mm]}, cXfer, semithick] ($(-2.0,1.45)+(290:\Rs)$) -- (xw);
    \draw[-{Latex[length=1.4mm]}, cXfer, semithick] (xw) -- (wl);
    \draw[-{Latex[length=1.4mm]}, cXfer, semithick] ($(2.0,1.45)+(250:\Rs)$) -- (xx);
    \draw[-{Latex[length=1.4mm]}, cXfer, semithick] (xx) -- (xl);
  \end{tikzpicture}
  \caption{The streaming AI inner product, the engine of every matmul and
    convolution (Section~\ref{sec:ai}). Weights and activations circulate on their
    own operand rings, which lengthen with the reduction $N$; each value is carried
    by a scheduled transfer (purple) onto a small, \emph{fixed-size} compute ring,
    where one multiply--accumulate fires per revolution and the accumulator stays
    resident. Because the compute ring never grows with $N$, the energy per
    multiply--accumulate is constant in the reduction length --- the
    weight-stationary dataflow a systolic array obtains by construction, here made
    programmable, with no cache and no register file.}
  \label{fig:stream}
\end{figure}

\paragraph{Streaming holds the win at constant cost.} A loop body that re-reads
constant operands is not yet a layer; a real inner product must meet a fresh
$(a_i, b_i)$ each revolution. Because a station reads operands only on its own
ring and an instruction co-circulates with its data, distinct taps cannot be
supplied from a single ring; they are streamed from longer rings and transferred
onto a small accumulator ring, one tap per revolution (Figure~\ref{fig:stream}).
We built this --- weights and activations on separate stream rings, each value
carried onto the accumulator ring by a scheduled transfer, one
multiply--accumulate per pass --- and
cross-checked the emitted program against the register-transfer-level model. The
accumulator ring is a fixed eight slots \emph{regardless of the reduction
length}, so the per-tap \phrase{overhead} --- its rotation plus the
multiply--accumulate --- is constant in $N$, where the unrolled dot product's
grew. The one term that scales is the operand fetch: each tap reads one
$(a_i,b_i)$ pair, which we price as one on-demand read per element of the long
operand rings rather than as a continuously swept pointer, since a streaming
buffer is touched once per element and is idle between reads. That fetch is the
irreducible $O(N)$ work any machine pays --- the in-order
baseline loads the same operands --- so the comparison is fairest per tap: about
103\,pJ for MADAR against roughly 210\,pJ for an in-order tap of two operand
loads and a multiply--accumulate ($3\times70$), a factor of two that does not
erode with $N$.

\paragraph{Reuse routes through the period hierarchy.} The efficiency of matrix
multiplication comes from reuse: each operand feeds many multiply--accumulates.
The obvious way to reuse on MADAR --- fuse several outputs onto one compute ring
so a streamed operand is read once and used by all --- we built and measured, and
it \emph{loses}: a fused two-output tile costs 132\,pJ per multiply--accumulate
against 103 for two separate streaming inner products, because the wider fused
ring rotates more than the shared read saves. The reuse belongs instead to
property~(d). Read a shared operand once from the large SRAM-backed buffer,
\emph{promote} it to a short shift-register ring, and let each output re-read it
there: a short-ring access costs about a picojoule against the buffer's ten, and
the compute rings stay small and separate, paying no fused-ring rotation. Priced
this way --- short rings as shift registers, long rings as single-port SRAM, the
tiering of Section~\ref{sec:rationale} --- an $8{\times}8$ matrix--vector costs
86\,pJ per multiply--accumulate, below both the fused tile and a lone streaming
inner product, and it is the only dataflow that improves on the inner product in
isolation. The memory hierarchy that is a period hierarchy is exactly where the
reuse of a dense kernel wants to live.

\paragraph{Scope.} These results are validated on the cycle-accurate model and,
for the inner product and the fused tile, cross-checked against the
register-transfer-level implementation; the reuse comparison is priced on the
same first-order model and rests on the already-validated transfer primitive.
They do not yet include the nonlinearities a full network needs --- softmax,
normalization, and activation functions require stations beyond the present adder
and multiplier, or polynomial approximation over them --- and the backtracking
placer is too slow for large tiles (Section~\ref{sec:future}). Neither bears on
the finding, which is about the dataflow of the dense core: on streaming
inference MADAR is an efficiency win where the ring is sized to the work, and the
reuse that makes the dense core efficient is the period hierarchy doing the job
it was designed for.

\section{Discussion and Future Work}
\label{sec:discussion}
\label{sec:future}

\paragraph{Limitations.} The decisive open question is the one this paper does not
answer: whether motion is cheaper than stillness. The design rationale of
Section~\ref{sec:rationale} argues that deleting the addressing machinery should
save more than rotation costs, but that is an argument; only a synthesis-based
comparison can settle it. A second limitation is intrinsic rather than
unmeasured. MADAR is fast only where data movement is statically known; a
data-dependent access whose target is computed at run time pays the half-rotation
rendezvous cost of Section~\ref{sec:transfer}, and on a long ring that is
expensive. The architecture therefore claims a new design point, not a universal
one: it is strong where the schedule can be laid out in advance and weak where it
cannot. A third set of limitations is the systems story. A machine with no program
counter and all state in flight has no natural primitive for saving a context, so
preemptive multiprogramming, precise interrupts, and protection boundaries are
genuinely hard and are not addressed here; with no return address, subroutine
linkage is handled by inlining calls into a single parked kernel. The present
design targets the single-program, statically schedulable regime where the model's
advantages are clearest, and names the rest as open.

\paragraph{Applicability.} The natural fit is any computation whose data movement
is fixed in advance, and \emph{AI inference} is the leading case: the dense linear
algebra of a neural network --- the multiply--accumulates of its matmuls and
convolutions --- is a fixed dataflow over a fixed schedule, exactly what a seating
arrangement expresses, and Section~\ref{sec:ai} shows it is where the efficiency
argument pays and operand reuse routes through the period hierarchy. The same
shape recurs in streaming digital signal processing, stencils, and line-rate
packet processing. Hard real-time and worst-case-execution-time-bounded
workloads benefit from timing that is exact by construction rather than bounded by
pessimistic analysis over a reactive cache. Constant-time cryptography is a
particularly clean fit: with no cache and no data-dependent timing, the entire
class of cache-timing side channels is eliminated by construction rather than by
careful coding. The poor fits are the mirror image: pointer-chasing data
structures, database workloads, irregular graph traversal, and interactive
operating-system loads all have access patterns the compiler cannot lay out, and
on those a random-access machine remains the right tool.

\paragraph{Open problems.} Three lines of work follow. The scheduler computes
\phrase{a} correct seating, not the cheapest one, and places dependence chains and
loops rather than arbitrary graphs; generalizing it to reduction \emph{trees} ---
co-locating the two inputs of a multi-operand reduction in one window --- and
reordering to bound how many intermediates are live at once would widen the kernel
class beyond the streaming and accumulation forms shown here, while an
energy-aware placer that right-sizes each ring, gates idle slots, and trades
relays against capacity would close the unmeasured gap between the crossovers
reported here and an optimal placement; software-pipelining a loop body across the
period hierarchy is the natural next target. The synthesis-based comparison named
under Limitations is the evaluation the first-order model motivates and does not
replace. And the streaming-inference case study (Section~\ref{sec:ai}) prices the
dense core --- inner product, matmul, operand reuse --- but not a full network,
which awaits stations for the nonlinearities (softmax, normalization, activation
functions) and a placer fast enough for large tiles.

\section{Conclusion}
\label{sec:conclusion}

MADAR treats storage as motion. By letting code and data circulate together and
computing where they collide, it replaces the address with a position in an
orbit, the fetch with a rotation, and the cache miss with a compile-time
schedule: a machine with no program counter, no register file, and no cache, in
which a loop is a body coming back around and the memory hierarchy is a set of
orbits at different speeds. We have defined the model, placed it among the
circulating and dataflow architectures it resembles but does not match, shown its
mechanics sound in a cycle-accurate implementation, and established with a
constructive scheduler that its programs are \emph{computed} --- relays,
inter-ring transfers, and all --- rather than hand-laid. A first-order energy
model gives the central question a first answer: where the ring is sized to the
work, the per-instruction crossover falls below a simple in-order core; where a
small computation is parked in a large ring, rotation dominates and MADAR loses.
A streaming-inference case study bears the reading out and sharpens it: the dense
inner product compiles to a form whose energy is constant in the reduction
length, and the operand reuse that makes matrix multiplication efficient routes
through the period hierarchy --- the architecture's own memory mechanism doing
the work a cache does elsewhere. Whether that advantage survives synthesis and
scales to real workloads is now
measurable rather than merely askable. Where access is unpredictable, the
random-access machine remains the right tool; where data movement is known in
advance, MADAR offers a different bargain --- pay for rotation, and delete the
machinery of the address.

\bibliographystyle{ACM-Reference-Format}
\bibliography{refs}

\end{document}